\documentclass{article}

\usepackage{microtype}
\usepackage{graphicx}
\usepackage{subcaption}
\usepackage{booktabs} %

\usepackage{hyperref}
\usepackage{titletoc}
\usepackage{xcolor}
\definecolor{codegray}{gray}{1.0}

\usepackage[accepted]{icml2026}

\usepackage{amsmath}
\usepackage{amssymb}
\usepackage{mathtools}
\usepackage{amsthm}

\usepackage[capitalize,noabbrev]{cleveref}

\usepackage[textsize=tiny]{todonotes}
\usepackage{graphicx}
\usepackage{tikz}
\usepackage{tikz-cd}
\usepackage{times}
\usepackage{courier}
\usepackage{bm}
\usepackage{physics}
\usepackage{xcolor}
\usepackage{natbib}
\usepackage{mdframed}
\usepackage{nicefrac}
\usepackage{booktabs}
\usepackage{lipsum}
\usepackage{titlesec}
\usepackage{wrapfig,lipsum,booktabs}
\usepackage{blindtext}
\usepackage{multirow}
\usepackage{subcaption}

\usepackage{algpseudocode}

\usepackage{titletoc}
\usepackage{todonotes}
\usepackage{setspace}
\usepackage{dsfont} %

\usepackage[capitalize,noabbrev]{cleveref}

\usepackage{CJK}

\usepackage{array}
\usepackage{makecell}

\def\({\left(}
\def\){\right)}
\def\[{\left[}
\def\]{\right]}

\usepackage{dashbox}
\usepackage{xcolor}
\usepackage{colortbl}
\definecolor{lightyellow}{rgb}{1.0, 0.95, 0.7}
\definecolor{Blue}{rgb}{0, 0, 0.8}
\definecolor{blue}{rgb}{0,0,1}
\definecolor{darkgreen}{rgb}{0,0.40,0}
\definecolor{firebrick}{rgb}{0.698,0.133,0.133}

\definecolor{colorA}{rgb}{1,0,0}
\definecolor{colorB}{rgb}{0,0.3,1}
\definecolor{colorC}{rgb}{0.9,0.8,0.2}
\definecolor{colorD}{rgb}{0,0.65,0}
\definecolor{lesslightgray}{rgb}{0.5,0.5,0.5}
\definecolor{light-gray}{gray}{0.95}

\let\cite\citep 

\usepackage{amsthm}
\makeatletter
\def\th@remark{%
  \thm@headfont{\bfseries}%
  \normalfont %
  \thm@preskip\topsep \divide\thm@preskip\tw@
  \thm@postskip\z@ %
}
\makeatother
\usepackage[many]{tcolorbox}
\theoremstyle{definition}

\tcolorboxenvironment{theorem}{
  breakable,
  colback=black!10,
  colframe=white,%
  width=\dimexpr\linewidth+10pt\relax,%
  enlarge left by=-5pt,%
  enlarge right by=-5pt,%
  boxsep=5pt,%
  boxrule=0pt,
  left=0pt,right=0pt,top=0pt,bottom=0pt,
  sharp corners,
  before skip=\topsep,
  after skip=\topsep
}

\tcolorboxenvironment{lemma}{
  breakable,
  colback=black!10,
  colframe=white,%
  width=\dimexpr\linewidth+10pt\relax,%
  enlarge left by=-5pt,%
  enlarge right by=-5pt,%
  boxsep=5pt,%
  boxrule=0pt,
  left=0pt,right=0pt,top=0pt,bottom=0pt,
  sharp corners,
  before skip=\topsep,
  after skip=\topsep
}

\tcolorboxenvironment{corollary}{
  breakable,
  colback=black!10,
  colframe=white,%
  width=\dimexpr\linewidth+10pt\relax,%
  enlarge left by=-5pt,%
  enlarge right by=-5pt,%
  boxsep=5pt,%
  boxrule=0pt,
  left=0pt,right=0pt,top=0pt,bottom=0pt,
  sharp corners,
  before skip=\topsep,
  after skip=\topsep
}

\theoremstyle{definition}

\tcolorboxenvironment{definition}{
  breakable,
  colback=black!10,
  colframe=white,%
  width=\dimexpr\linewidth+10pt\relax,%
  enlarge left by=-5pt,%
  enlarge right by=-5pt,%
  boxsep=5pt,%
  boxrule=0pt,
  left=0pt,right=0pt,top=0pt,bottom=0pt,
  sharp corners,
  before skip=\topsep,
  after skip=\topsep
}

\tcolorboxenvironment{assumption}{
  breakable,
  colback=black!10,
  colframe=white,%
  width=\dimexpr\linewidth+10pt\relax,%
  enlarge left by=-5pt,%
  enlarge right by=-5pt,%
  boxsep=5pt,%
  boxrule=0pt,
  left=0pt,right=0pt,top=0pt,bottom=0pt,
  sharp corners,
  before skip=\topsep,
  after skip=\topsep
}

\tcolorboxenvironment{problem}{
  breakable,
  colback=black!10,
  colframe=white,%
  width=\dimexpr\linewidth+10pt\relax,%
  enlarge left by=-5pt,%
  enlarge right by=-5pt,%
  boxsep=5pt,%
  boxrule=0pt,
  left=0pt,right=0pt,top=0pt,bottom=0pt,
  sharp corners,
  before skip=\topsep,
  after skip=\topsep
}

\crefname{theorem}{Theorem}{Theorems}
\crefname{proposition}{Proposition}{Propositions}
\crefname{lemma}{Lemma}{Lemmas}
\crefname{corollary}{Corollary}{Corollaries}
\crefname{definition}{Definition}{Definitions}
\crefname{assumption}{Assumption}{Assumptions}
\crefname{remark}{Remark}{Remarks}
\crefname{problem}{Problem}{Problems}
\crefname{property}{Property}{property}

\tcolorboxenvironment{hypothesis}{
  breakable,
  colback=black!10,
  colframe=white,%
  width=\dimexpr\linewidth+10pt\relax,%
  enlarge left by=-5pt,%
  enlarge right by=-5pt,%
  boxsep=5pt,%
  boxrule=0pt,
  left=0pt,right=0pt,top=0pt,bottom=0pt,
  sharp corners,
  before skip=\topsep,
  after skip=\topsep
}
\crefname{hypothesis}{Hypothesis}{Hypothesises}

\tcolorboxenvironment{fact}{
  breakable,
  colback=black!10,
  colframe=white,%
  width=\dimexpr\linewidth+10pt\relax,%
  enlarge left by=-5pt,%
  enlarge right by=-5pt,%
  boxsep=5pt,%
  boxrule=0pt,
  left=0pt,right=0pt,top=0pt,bottom=0pt,
  sharp corners,
  before skip=\topsep,
  after skip=\topsep
}
\crefname{fact}{Fact}{Facts}

\tcolorboxenvironment{example}{
  breakable,
  colback=black!10,
  colframe=white,%
  width=\dimexpr\linewidth+10pt\relax,%
  enlarge left by=-5pt,%
  enlarge right by=-5pt,%
  boxsep=5pt,%
  boxrule=0pt,
  left=0pt,right=0pt,top=0pt,bottom=0pt,
  sharp corners,
  before skip=\topsep,
  after skip=\topsep
}
\crefname{example}{Example}{Examples}

\tcolorboxenvironment{question}{
  breakable,
  colback=black!10,
  colframe=white,%
  width=\dimexpr\linewidth+10pt\relax,%
  enlarge left by=-5pt,%
  enlarge right by=-5pt,%
  boxsep=5pt,%
  boxrule=0pt,
  left=0pt,right=0pt,top=0pt,bottom=0pt,
  sharp corners,
  before skip=\topsep,
  after skip=\topsep
}
\crefname{question}{Question}{Questions}

\numberwithin{equation}{section}
\numberwithin{theorem}{section}
\numberwithin{proposition}{section}
\numberwithin{definition}{section}
\numberwithin{lemma}{section}
\numberwithin{assumption}{section}
\numberwithin{remark}{section}

\usepackage{lipsum}

\makeatletter
\let\save@mathaccent\mathaccent
\newcommand*\if@single[3]{%
    \setbox0\hbox{${\mathaccent"0362{#1}}^H$}%
    \setbox2\hbox{${\mathaccent"0362{\kern0pt#1}}^H$}%
    \ifdim\ht0=\ht2 #3\else #2\fi
}
\newcommand*\rel@kern[1]{\kern#1\dimexpr\macc@kerna}
\newcommand*\widebar[1]{\@ifnextchar^{{\wide@bar{#1}{0}}}{\wide@bar{#1}{1}}}
\newcommand*\wide@bar[2]{\if@single{#1}{\wide@bar@{#1}{#2}{1}}{\wide@bar@{#1}{#2}{2}}}
\newcommand*\wide@bar@[3]{%
    \begingroup
    \def\mathaccent##1##2{%
        \let\mathaccent\save@mathaccent
        \if#32 \let\macc@nucleus\first@char \fi
        \setbox\z@\hbox{$\macc@style{\macc@nucleus}_{}$}%
        \setbox\tw@\hbox{$\macc@style{\macc@nucleus}{}_{}$}%
        \dimen@\wd\tw@
        \advance\dimen@-\wd\z@
        \divide\dimen@ 3
        \@tempdima\wd\tw@
        \advance\@tempdima-\scriptspace
        \divide\@tempdima 10
        \advance\dimen@-\@tempdima
        \ifdim\dimen@>\z@ \dimen@0pt\fi
        \rel@kern{0.6}\kern-\dimen@
        \if#31
        \overline{\rel@kern{-0.6}\kern\dimen@\macc@nucleus\rel@kern{0.4}\kern\dimen@}%
        \advance\dimen@0.4\dimexpr\macc@kerna
        \let\final@kern#2%
        \ifdim\dimen@<\z@ \let\final@kern1\fi
        \if\final@kern1 \kern-\dimen@\fi
        \else
        \overline{\rel@kern{-0.6}\kern\dimen@#1}%
        \fi
    }%
    \macc@depth\@ne
    \let\math@bgroup\@empty \let\math@egroup\macc@set@skewchar
    \mathsurround\z@ \frozen@everymath{\mathgroup\macc@group\relax}%
    \macc@set@skewchar\relax
    \let\mathaccentV\macc@nested@a
    \if#31
    \macc@nested@a\relax111{#1}%
    \else
    \def\gobble@till@marker##1\endmarker{}%
    \futurelet\first@char\gobble@till@marker#1\endmarker
    \ifcat\noexpand\first@char A\else
    \def\first@char{}%
    \fi
    \macc@nested@a\relax111{\first@char}%
    \fi
    \endgroup
    }
\makeatother

\makeatletter
\newcommand*{\redefinesymbolwitharg}[1]{%
  \expandafter\let\csname ltx#1\expandafter\endcsname\csname #1\endcsname
  \@namedef{#1}{\@ifnextchar{^}{\@nameuse{#1@}}{\@nameuse{#1@}^{}}}%
  \expandafter\def\csname #1@\endcsname^##1##2{%
     \csname ltx#1\endcsname\ifx!##1!\else^{##1}\fi\mathopen{}\mathclose\bgroup\left(##2\aftergroup\egroup\right)
     }%
}
\makeatother
\redefinesymbolwitharg{sin}
\redefinesymbolwitharg{cos}

\usepackage{listings}
\usepackage{xcolor}

\definecolor{codegray}{gray}{0.95}
\definecolor{codered}{rgb}{0.6,0,0}
\definecolor{codeblue}{rgb}{0,0,0.6}
\definecolor{codegreen}{rgb}{0,0.5,0}

\lstdefinestyle{pythonstyle}{
    backgroundcolor=\color{codegray},
    commentstyle=\color{codegreen},
    keywordstyle=\color{codered}\bfseries,
    numberstyle=\tiny\color{gray},
    stringstyle=\color{codeblue},
    basicstyle=\ttfamily\small,
    breaklines=true,
    captionpos=b,
    keepspaces=true,
    numbers=left,
    numbersep=5pt,
    showspaces=false,
    showstringspaces=false,
    showtabs=false,
    tabsize=4,
    language=Python
}
\newcommand{\methodac}{G\textsc{enome}-F\textsc{actory}}

\icmltitlerunning{{\methodac: A Library for Tuning, Deploying, and Interpreting Genomic Foundation Models}}

\begin{document}

\twocolumn[
\icmltitle{{\methodac: A Library for Tuning, Deploying, and Interpreting Genomic Foundation Models}}

  \icmlsetsymbol{equal}{*}

  \begin{icmlauthorlist}
    \icmlauthor{Weimin Wu}{equal,nu_cs}
    \icmlauthor{Xuefeng Song}{equal,nu_cs}
    \icmlauthor{Yibo Wen}{equal,nu_cs}\\
    \icmlauthor{Qinjie Lin}{nu_cs} 
    \icmlauthor{Zhihan Zhou}{nu_cs}
    \icmlauthor{Jerry Yao-Chieh Hu}{nu_cs}
    \icmlauthor{Zhong Wang}{um_ns,lbnl,lbnl_gs}
    \icmlauthor{Han Liu}{nu_cs,nu_ds}
  \end{icmlauthorlist}

  \icmlaffiliation{nu_cs}{Center for Foundation Models and Generative AI \& Department of Computer Science, Northwestern University, USA}
  \icmlaffiliation{um_ns}{School of Natural Sciences, University of California at Merced, USA}
  \icmlaffiliation{lbnl}{Department of Energy Joint Genome Institute, Lawrence Berkeley National Laboratory, USA}
  \icmlaffiliation{lbnl_gs}{Environmental Genomics and Systems Biology Division, Lawrence Berkeley National Laboratory, USA}
  \icmlaffiliation{nu_ds}{Department of Statistics and Data Science, Northwestern University, USA}

  \icmlcorrespondingauthor{Weimin Wu}{wwm@u.northwestern.edu}
  \icmlcorrespondingauthor{Han Liu}{hanliu@northwestern.edu}

  \icmlkeywords{Machine Learning, ICML}

  \vskip 0.3in
]

\printAffiliationsAndNotice{\icmlEqualContribution}

\begin{abstract}
We introduce \methodac, the first integrated Python library for tuning, deploying, and interpreting genomic foundation models.
Our core contribution is to simplify and unify the workflow for genomic model development: data collection, model tuning, inference, benchmarking, and interpretability.
For data collection, \methodac~offers an automated pipeline to download genomic sequences and preprocess them. 
For model tuning, \methodac~supports both full and parameter-efficient fine-tuning across diverse genomic models.
For inference, \methodac~enables both embedding extraction and DNA sequence generation.
For benchmarking, we include two existing benchmarks and provide a flexible interface to incorporate additional benchmarks.
For interpretability, \methodac~introduces an open-source biological interpreter based on a sparse auto-encoder. 
We validate the utility of \methodac~across three dimensions:
(i) Compatibility with diverse models and fine-tuning methods;
(ii) Benchmarking downstream performance using two open-source benchmarks;
(iii) Biological interpretation of learned representations with DNABERT-2.
These results highlight its practical value for real-world genomic analysis.
\noindent\textbf{GitHub}: \url{https://github.com/WeiminWu2000/Genome_Factory}.
\end{abstract}

\titlespacing*{\paragraph}
{0pt}        %
{0.30ex}      %
{0.63em}      %

\section{Introduction}
\label{sec:intro}

We introduce \methodac, the first integrated Python library for tuning, deploying, and interpreting genomic foundation models (GFMs).
GFMs have transformed biological research by enabling accurate and scalable solutions to key tasks, including epigenetic prediction \cite{Gao2024-ns}, regulatory element discovery \cite{Hwang2024-jh}, and species classification \cite{zhou2025dnabert}.
These models learn from large-scale genomic data and support progress in personalized medicine, evolutionary biology, and functional genomics \cite{Consens2025-sl}.
Despite the potential of GFMs, their adoption in life sciences remains limited due to a fundamental gap between domain expertise and technical implementation. 
On one hand, engineers handle model training and deployment but often lack a biological context.
Conversely, biologists design experiments and define scientific goals but lack expertise for large models.
To address this, \methodac~offers a unified, practical platform to bridge this gap and accelerate the broader use of GFMs in life science research.

While general-purpose language model fine-tuning frameworks such as LLaMA-Factory \cite{zheng2024llamafactory} provide integrated tools, they do not address the unique requirements of genomics.
Firstly, genomic data demands specialized handling, including support for domain-specific formats like FASTA, integration with repositories such as NCBI \cite{geer2010ncbi} for data acquisition, and domain-specific data preprocessing.
Secondly, developers have built genomic models across diverse environments with heterogeneous dependencies.
This lack of standardization makes it difficult to use models within a unified framework, and even harder to ensure compatibility with tools from the language model ecosystem.
Thirdly, fine-tuning genomic models differs from language objectives: rather than instruction tuning or text generation, biological tasks often involve predicting variant effects, enhancer activity, or gene expression.
These tasks require custom model adaptations and biology-informed loss functions aligned with real-world genomic objectives.
Fourthly, evaluation further depends on domain-specific benchmarks, such as variant detection or regulatory site classification \cite{zhou2024dnabert}.
These diverge from the text-based benchmarks used to assess language models.
Finally, biological interpretability is central to the utility of GFMs for scientists, whereas it is not a focus of existing language model fine-tuning frameworks.
As a result, the GFMs field still lacks a unified, user-friendly platform to support the full pipeline for tuning and deploying models.

To address this challenge, we introduce \methodac, the first unified Python library for fine-tuning, deploying, and interpreting genomic models. 
\methodac~features six modular components.
(i) \textbf{Genome Collector}: Retrieves genomic sequences from public repositories (e.g., NCBI \cite{geer2010ncbi}) and applies preprocessing such as GC content normalization and N nucleotide correction. 
It also includes task-specific dataset builders for histone modification, enhancer, and promoter classification, with automated region extraction and labeling.
(ii) \textbf{Model Loader}: Supports diverse GFMs, including discriminative models—HyenaDNA \cite{nguyen2023hyenadna}, DNABERT-2 \cite{zhou2024dnabert}, Caduceus \cite{schiff2024caduceus}, and Nucleotide Transformer \cite{dalla2025nucleotide}—as well as generative models such as EVO \cite{nguyen2024sequence} and GenomeOcean \cite{zhou2025genomeocean}.
(iii) \textbf{Model Trainer}: Enables full-parameter fine-tuning and parameter-efficient methods such as low-rank adaptation (LoRA) \cite{hu2022lora} and adapter tuning \cite{he2021effectiveness}.
It applies to both classification and regression tasks.
(iv) \textbf{Inference Engine}: Facilitates both embedding extraction and sequence generation.
(v) \textbf{Benchmarker}: Provides two built-in, open-source genomic benchmarks and a plugin system for incorporating custom, domain-specific evaluation tasks and datasets.
(vi) \textbf{Biological Interpreter}: Enhances model interpretability through a sparse auto
encoder. 
It disentangles embeddings into sparse latent units and aligns them with genomic features via regression against external biological readouts. 
This provides an open-source tool to interpret the internal representations of GFMs.
\methodac~also offers user-friendly interfaces: a zero-code command-line interface (CLI) and an intuitive Gradio-based web-based user interface (WebUI) \cite{abid2019gradio}.
These support both non-expert users and advanced developers with minimal computational and engineering effort.

In summary, we have the following three main contributions:
\begin{itemize}
    \item We introduce \methodac, the first integrated Python framework to streamline the complete genomic model workflow. 
    It integrates six components: 
    (i) Genome Collector: comprehensive data collection and preprocessing (\cref{subsec:collector});
    (ii) Model Loader: broad support for diverse genomic models (\cref{subsec:loader});
    (iii) Model Trainer: an easy-to-use fine-tuning pipeline (\cref{subsec:trainer});
    (iv) Inference Engine: efficient embedding extraction and sequence generation (\cref{subsec:inference});
    (v) Benchmarker: built-in genomic benchmarks and extensible evaluation plugins (\cref{subsec:benchmark}); 
    (vi) Biological Interpreter: model interpretability via sparse auto-encoder (\cref{subsec:Interpreter}).
    
    \item Beyond flexibility and ease of use, \methodac~is the first unified framework to bring together diverse genomic models under a single interface.
    This enables seamless model comparison and assists users in selecting the most suitable model for a customized task.  
    Notably, the Biological Interpreter is an open-source tool to decode the internal representations of GFMs with a sparse auto-encoder.
    This provides deeper biological insights into models.

    \item We validate the utility of \methodac~across three dimensions:
    (i) Compatibility with diverse genomic foundation models and three widely used fine-tuning methods;
    (ii) Benchmarking downstream performance using two open-source benchmarks: Genome Understanding Evaluation (GUE) Benchmark \cite{zhou2024dnabert} and Genomic Benchmarks \cite{grevsova2023genomic};
    (iii) Biological interpretation of learned representations with DNABERT-2.
    These results clearly highlight its end-to-end usability and practical value for real-world genomic analysis.

\end{itemize}

\paragraph{Organization.}
\cref{sec:framework} details the \methodac, including Genome Collector, Model Loader, Model Trainer, Inference Engine, Benchmarker, and Biological Interpreter.
\cref{sec:Empirical Study} presents the results of our experiments to evaluate \methodac’s effectiveness. 
We leave related work in \cref{sec:related work}, to discuss genomic foundation models and libraries for language models.

\paragraph{Conflict of Interest Disclosure.}
The authors declare no financial conflicts of interest. The evaluated models are open-source or publicly available research models, and no author is employed by a company whose proprietary model is evaluated in this study. Funding sources and institutional support are disclosed in the Acknowledgements.

\section{{\methodac}~Framework}
\label{sec:framework}

\begin{figure*}[t]
    \centering 
    \includegraphics[width=0.98\textwidth]{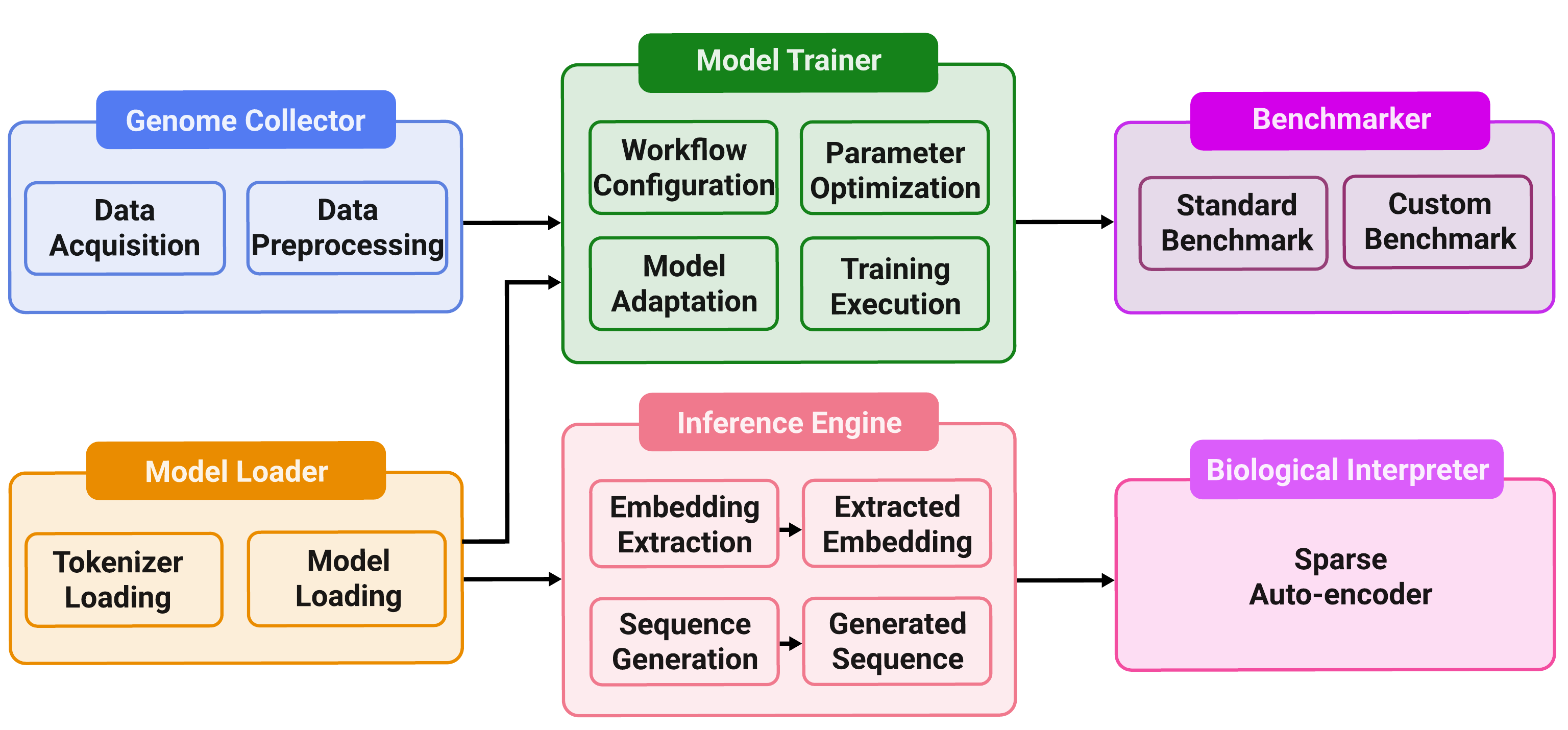} 
    \caption{\textbf{Overview of \methodac.} 
    The framework consists of six components. 
    \textbf{Genome Collector} acquires genomic sequences from public repositories and performs preprocessing (e.g., GC normalization). 
    \textbf{Model Loader} supports major genomic models (e.g., HyenaDNA, DNABERT-2, Caduceus, Nucleotide Transformer, EVO and GenomeOcean) and their tokenizers. 
    \textbf{Model Trainer} configures workflows, adapts models to classification or regression tasks, and executes training with full fine-tuning or parameter-efficient methods (LoRA, adapters). 
    \textbf{Inference Engine} enables embedding extraction and sequence generation. 
    \textbf{Benchmarker} provides standard benchmarks and allows integration of custom tasks. 
    \textbf{Biological Interpreter} enhances interpretability through sparse auto-encoders.}
    \label{fig:overview} 
\end{figure*}

\methodac~comprises six core modules to unify the workflow for genomic models.
The Genome Collector (\cref{subsec:collector}) simplifies data retrieval (e.g., from NCBI \cite{geer2010ncbi}) and integrates preprocessing pipelines.
The preprocessing includes the quality control steps, such as sequence-length filtering, GC content normalization, and correction of N nucleotide.
It also supports task-specific dataset builders for histone modification, enhancer, and promoter classification, with automated region extraction, chromosome-name harmonization, and labeling.
Model Loader (\cref{subsec:loader}) and Model Trainer (\cref{subsec:trainer}) handle model loading and fine-tuning.
They support both full and parameter-efficient methods, such as LoRA and adapters for classification and regression tasks.
The Inference Engine (\cref{subsec:inference}) enables embedding extraction and sequence generation with pre-trained genomic foundation models.
The Benchmarker (\cref{subsec:benchmark}) provides essential tools and datasets for rigorous model evaluation and comparison across tasks.
The Biological Interpreter (\cref{subsec:Interpreter}) delivers interpretability via a sparse auto-encoder.
It learns near-monosemantic latent units from model embeddings and links them to interpretable genomic features (e.g., motif presence, sequence length) through regression on external biological readouts.
Users can access \methodac~through a zero-code command-line interface or an interactive web-based user interface (\cref{subsec:webui}).
This design supports both flexibility and scalability for diverse applications. 
We provide an overview of \methodac~in \cref{fig:overview}, and {\bf detailed usage examples} in \cref{app_sec:use}.

\subsection{Genome Collector}
\label{subsec:collector}
The Genome Collector manages the upstream data pipeline for genomic models. 
It automates the fetching, transformation, and validation of genomic data.
This extends beyond basic sequence downloads to include regulatory and epigenetic annotations.
Standardizing dataset construction across diverse biological tasks, it ensures high-quality, task-ready inputs for the model.
Furthermore, the framework supports custom multi-stage bioinformatics pipelines.

\paragraph{Data Acquisition.}  
The Genome Collector offers multiple flexible pipelines for downloading and preparing diverse DNA sequence datasets. 
Beyond basic genome-wide retrieval from public repositories such as NCBI, the framework supports three task-driven acquisition modes. 
(i) It constructs region-based datasets by identifying signal-enriched locations, such as high-coverage intervals from genome-wide profiles (e.g., histone modification enrichment). 
(ii) It builds binary classification datasets by separating annotated functional regions from background sequences, such as distinguishing regulatory elements from random regions (e.g., enhancers). 
(iii) It samples region-versus-background pairs by comparing annotated start sites to non-start regions of matched length (e.g., promoters). 
Each pipeline handles file downloading, genome indexing, sequence extraction, chromosome name harmonization, and sequence quality filtering. 
A unified interface lets users choose the task type and generate corresponding datasets with minimal effort.

\paragraph{Data Preprocessing.}  
Each acquisition mode applies task-specific parsing and transformation logic.
For species-level classification, the system samples fixed-length DNA fragments from genomes and assigns species labels.
The histone modification pipeline extracts gene-body sequences aligned to signal peaks and binarizes them into high/low classes based on enrichment thresholds.
The enhancer and promoter pipelines define positive regions from regulatory annotations and sample negatives from non-overlapping regions.
Genome Collector saves each dataset in a standardized format with DNA sequences and corresponding labels, and partitions it into training, validation, and testing sets.

\paragraph{Quality Control and Data Cleaning.} 
To ensure reliable and robust downstream performance, Genome Collector applies a multi-stage quality control protocol.
Initial filters enforce several basic constraints on sequence length, GC content, and the overall proportion of ambiguous nucleotides.
Further statistical quality control removes outliers using three strategies:
(i) filtering sequences with excessive N nucleotide;
(ii) applying chi-square tests to detect compositional biases relative to expected nucleotide distributions;
and (iii) removing rare compositional profiles below a predefined frequency threshold.
Quality control steps confirm that the cleaned datasets exhibit balanced GC content and sequence length distributions.
These safeguards ensure the final datasets have valid biological structure and robust statistical properties for training genomic models.

\subsection{Model Loader}
\label{subsec:loader}
The Model Loader handles the initial phase of inference or fine-tuning by loading the genomic model and corresponding tokenizer.
It leverages the HuggingFace Transformers \cite{wolf2019huggingface} to support diverse models, including HyenaDNA \cite{nguyen2023hyenadna}, DNABERT-2 \cite{zhou2024dnabert}, Caduceus \cite{schiff2024caduceus}, Nucleotide Transformer \cite{dalla2025nucleotide}), EVO \cite{nguyen2024sequence} and GenomeOcean \cite{zhou2025genomeocean}.

\paragraph{Tokenizer Loader.}
The system automatically loads the appropriate tokenizer for the selected genomic model using the Hugging Face tokenizer API.
This ensures accurate and consistent encoding of input DNA sequences and full compatibility with the model’s input format.

\paragraph{Model Architecture and Checkpoint Loader.} 
After initializing the tokenizer, the system loads the model with either pretrained checkpoints or random parameters.
It configures the architecture based on the selected fine-tuning method.
For full-parameter or LoRA fine-tuning, it loads the model and attaches a task-specific classification or regression head.
For adapter fine-tuning, it freezes the base model’s weights and inserts an external adapter module for task-specific training.

\subsection{Model Trainer}
\label{subsec:trainer}
Adapting large genomic models to downstream tasks poses significant computational challenges due to their size and parameter count.
The Model Trainer manages configuration, launches fine-tuning jobs, and monitors training.
Furthermore, the framework supports the joint optimization of data preprocessing steps, such as GC normalization, together with model training, and also supports multi-task learning.
The module includes four components: workflow configuration, parameter optimization, model adaptation, and training execution.
\cref{tab:techniques-comparison} summarizes the training strategies and system-level optimizations.

\begin{table}[htbp]
  \centering
  \caption{{\bf Featured tuning techniques.}}
  \setlength{\tabcolsep}{4pt}
  \begin{tabular}{lccc}
    \toprule
                       & \textbf{Full} & \textbf{LoRA} & \textbf{Adapter} \\
    \midrule
    \textbf{Mixed precision}       & $\checkmark$ & $\checkmark$ & $\checkmark$ \\
    \textbf{Flash attention}       & $\checkmark$ & $\checkmark$ & $\checkmark$ \\
    \textbf{Gradient accumulation} & $\checkmark$ & $\checkmark$ & $\checkmark$ \\
    \bottomrule
  \end{tabular}
  \label{tab:techniques-comparison}
\end{table}

\paragraph{Workflow Configuration.}
The system automatically reads user inputs and builds task-specific configurations for a classification or regression task.
Custom parsers validate hyperparameters and construct training pipelines based on the selected fine-tuning strategy.
The Hugging Face Trainer manages core training and distributed execution to ensure reliability, consistency, and scalability.

\paragraph{Parameter Optimization.}
\methodac~offers three fine-tuning strategies to effectively balance resource efficiency and model adaptability:
(i) Full-parameter fine-tuning updates all model parameters.
It maximizes task adaptation but incurs a high computational cost.
(ii) Low-rank adaptation (LoRA) \cite{hu2022lora} freezes the base model and introduces trainable low-rank matrices in attention or feed-forward layers.
It reduces memory and training time while preserving performance.
(iii) Adapter tuning \cite{he2021effectiveness} adds a lightweight neural module (e.g., multilayer perceptrons \cite{popescu2009multilayer} or convolutional neural network \cite{o2015introduction}) after the frozen base model.
It only updates the adapter parameters.
All methods support customizable hyperparameters, e.g., learning rate, weight decay, LoRA rank, and scaling factor.

\paragraph{Model Adaptation.} 
\methodac~adjusts model architectures based on the task.
For classification, it appends activation functions such as Softmax to the output layer and applies cross-entropy loss \cite{de2005tutorial}.
For regression, it sets up continuous-valued outputs and uses mean squared error loss \cite{schluchter2005mean}.
The system handles variable-length sequences using dynamic padding or truncation to maintain compatibility with input requirements.

\paragraph{Training Execution.} 
\methodac~ seamlessly integrates the following performance optimizations:
(i) Precision control supports full, FP16, and BF16 formats to reduce memory usage and speed up training on compatible hardware.
(ii) Flash attention \cite{dao2022flashattention} accelerates attention computation and minimizes memory by avoiding explicit intermediate matrices.
(iii) Gradient accumulation simulates large batch sizes, and learning rate scheduling improves convergence stability.
It also applies gradient clipping to prevent exploding gradients.
Furthermore, it supports multi-GPU training via distributed data parallel \cite{li2020pytorch}.
During training, it logs evaluation metrics at configurable intervals.
For classification tasks, it tracks accuracy, F1 score, precision, recall, and Matthews correlation coefficient. 
For regression, it records mean squared error and mean absolute error.
The system saves checkpoints and retains the best-performing one based on validation metrics.
This enables flexible training resumption.

\subsection{Inference Engine}
\label{subsec:inference}

The Inference Engine offers a unified interface for applying genomic foundation models to two key tasks: sequence embedding extraction and DNA sequence generation.

\paragraph{Embedding Extraction.} 
This component processes input DNA sequences, runs the model in evaluation mode, and extracts the final hidden state as the sequence embedding.
These embeddings support downstream tasks such as classification, regression, clustering, and visualization.

\paragraph{Sequence Generation.}
This component enables compatible models to flexibly generate diverse novel DNA sequences from user-provided prompts.
It supports applications such as in silico sequence variation, synthetic data augmentation, and functional sequence exploration.

\subsection{Benchmarker}
\label{subsec:benchmark}

The Benchmarker module provides essential tools for evaluating genomic foundation models on both classification and regression tasks.
It supports standardized benchmark integration, flexible plugins for custom domain-specific evaluation tasks, and automated performance evaluation.

\paragraph{Incorporating Benchmarks.}
\methodac~includes two benchmark suites: the Genome Understanding Evaluation (GUE) Benchmark \cite{zhou2024dnabert} and Genomic Benchmarks \cite{grevsova2023genomic}.
It also supports plugins for integrating custom, domain-specific evaluation tasks.
To use a custom benchmark, users format their data according to the Model Trainer's input schema.
For classification, the dataset should include three CSV files (training, validation, and testing).
Each contains two columns: one for DNA sequences and one for integer labels.
For regression, users follow the same structure but replace the label column with continuous values.

\paragraph{Evaluating Models.}
\methodac~runs the selected model on the benchmark dataset and records detailed task-specific metrics.
For classification, it computes accuracy, F1 score, precision, recall, and Matthews correlation coefficient.
For regression, it computes mean squared error and mean absolute error.
The system logs all results in a structured JSON file.
This allows users to easily compare model performance across tasks, datasets, or different training strategies.

\subsection{Biological Interpreter}
\label{subsec:Interpreter}
The Biological Interpreter module enables meaningful interpretation of genomic foundation models by explicitly linking internal model representations to biological features. 
This supports novel hypothesis generation and provides deeper insight into what the model has actually learned.
However, it should be viewed as an initial step toward mechanistic interpretability for GFMs using sparse autoencoders, rather than a fully validated tool for biological discovery.

\paragraph{Sparse Auto-encoder.}  
\methodac~uses a sparse auto-encoder to disentangle latent embeddings from genomic models. 
The workflow involves three stages:
(i) Sequence embedding extraction: 
The system embeds input DNA sequences with a pretrained genomic model, such as GenomeOcean.
(ii) Sparse auto-encoder training: 
It trains a sparse auto-encoder on these embeddings with a reconstruction loss.
The training enforces sparsity so that only a small subset of latent units activate per sequence. 
This encourages each unit to capture a distinct, near-monosemantic genomic feature.
(iii) Regression to external readouts: 
The system fits regression models between the sparse latent units and external biological readouts (e.g., sequence length, motif presence, or experimental measurements) to associate individual neurons with interpretable molecular features.

\subsection{Command-line and Web-based User Interface}
\label{subsec:webui}

\methodac~offers both a command-line interface (CLI) and a web-based user interface (WebUI) to accommodate a range of user preferences and expertise levels.

\paragraph{Command-line Interface.}
Users access the command-line interface through a single convenient entry point, \texttt{genomefactory-cli}, and define configuration-first workflows with YAML.
Users specify tasks, datasets, models, and training or inference settings through compact configuration files.
The command-line interface supports four core functionalities: 
(i) data acquisition and preprocessing with task-specific dataset builders and automated quality control; 
(ii) model training using full fine-tuning or parameter-efficient methods; 
(iii) inference for embedding extraction and sequence generation; 
and (iv) interpretability using the sparse auto-encoder–based Biological Interpreter.
The system saves all metrics and artifacts for downstream evaluation and reproducibility.

\paragraph{Web-based User Interface.}
The WebUI uses Gradio \cite{abid2019gradio} to complement the CLI and gives users an intuitive, code-free way to conveniently access all core \methodac~features.
Users configure data, models, and tasks for training, inference, benchmarking, and interpretability through a clear graphical layout.
The interface shows relevant parameters based on the selected method, applies sensible defaults, and organizes workflows into task-specific tabs.
Users launch tasks with a single click and view logs and results in real time within the browser.
By hiding the underlying code, the WebUI lets researchers interact with models through a graphical interface.

\section{Empirical Study}
\label{sec:Empirical Study}We comprehensively demonstrate the effectiveness of \methodac~through three key dimensions.
Each dimension highlights the partial capabilities of six integrated modules:
(i) Fine-tuning compatibility across diverse models (\cref{subsec:downstream}):
This dimension evaluates the compatibility of three fine-tuning strategies: full, LoRA, and adapter tuning.
It covers a range of models. 
It showcases the functionality of the Model Loader, Model Trainer, and Inference Engine.
(ii) Benchmarking downstream performance (\cref{subsec:Comparison}):
By benchmarking different models on standardized downstream tasks, we analyze their trade-offs between accuracy and computational efficiency. 
This dimension emphasizes the role of the Benchmarker module.
(iii) Biological interpretation of learned representations with DNABERT-2 (\cref{subsec:bio_inter}):
We explore how model embeddings capture biological signals, such as correlations with sequence length, to assess interpretability. 
This dimension demonstrates the capability of the Genome Collector and Biological Interpreter modules.

\subsection{Fine-tuning of Diverse Models}
\label{subsec:downstream}
We evaluate the six genomic foundation models from the Model Loader using all three fine-tuning strategies from the Model Trainer. 
Notably, the adapter-based method requires extracting base model embeddings before training.
This highlights the functionality of the Inference Engine.

\begin{table*}[t]
  \centering 
    \caption{\textbf{Fine-tuning efficiency with different methods in \methodac.} 
    We report the number of trainable parameters, peak GPU memory usage, and throughput (thousands of tokens per second). 
    “---” marks settings we did not evaluate due to computational constraints (e.g., EVO with full fine-tuning). 
    We conduct all experiments on a single NVIDIA H100 (80GB). 
    Due to the large size of EVO (7B) and its computational demands, we exclude full-parameter fine-tuning for this model. 
    "Full" denotes full-parameter fine-tuning, "LoRA" denotes low-rank adaptation, and "Adapter" denotes adapter-based tuning.}
  \label{tab:training-efficiency-part1}
  \resizebox{1.0\textwidth}{!}{
    \begin{tabular}{l*{3}{ccc}}
      \toprule
      & \multicolumn{3}{c}{\textbf{HyenaDNA-160k}}
      & \multicolumn{3}{c}{\textbf{DNABERT2}}
      & \multicolumn{3}{c}{\textbf{Caduceus-131k}} \\
      \cmidrule(lr){2-4}
      \cmidrule(lr){5-7}
      \cmidrule(lr){8-10}
      \textbf{Method}
      & \textbf{Trainable} & \textbf{Mem} & \textbf{Throughput}
      & \textbf{Trainable} & \textbf{Mem} & \textbf{Throughput}
      & \textbf{Trainable} & \textbf{Mem} & \textbf{Throughput} \\
      & \textbf{params} & \textbf{(GB)} & \textbf{(KTok/s)}
      & \textbf{params} & \textbf{(GB)} & \textbf{(KTok/s)}
      & \textbf{params} & \textbf{(GB)} & \textbf{(KTok/s)} \\
      \midrule
      \textbf{Full}
      &   6.55M &  5.75 & 381.48
      & 117.08M &  7.30 &  40.70
      &   7.73M &  6.89 & 143.55 \\
      \textbf{LoRA}
      &   0.10M &  4.45 & 446.05
      &   1.49M &  6.18 &  44.07
      &   0.81M &  1.85 & 394.30 \\
      \textbf{Adapter}
      &   0.07M &  1.84 & 1113.32
      &   0.20M &  2.09 & 124.30
      &   0.07M &  1.85 & 467.72 \\
      \bottomrule
    \end{tabular}
  } 

    \bigskip

  \resizebox{1.0\textwidth}{!}{
    \begin{tabular}{l *{3}{ccc}}
  \toprule
  & \multicolumn{3}{c}{\textbf{Nucleotide Transformer-500M}}
  & \multicolumn{3}{c}{\textbf{EVO-1-131k}}
  & \multicolumn{3}{c}{\textbf{GenomeOcean-100M}} \\
  \cmidrule(lr){2-4} \cmidrule(lr){5-7} \cmidrule(lr){8-10}
  \textbf{Method}
    & \textbf{Trainable} & \textbf{Mem} & \textbf{Throughput}
    & \textbf{Trainable} & \textbf{Mem} & \textbf{Throughput}
    & \textbf{Trainable} & \textbf{Mem} & \textbf{Throughput} \\
  & \textbf{params} & \textbf{(GB)} & \textbf{(KTok/s)}
    & \textbf{params} & \textbf{(GB)} & \textbf{(KTok/s)}
    & \textbf{params} & \textbf{(GB)} & \textbf{(KTok/s)} \\
  \midrule
  \textbf{Full}
    & 480.45M & 18.94 &  12.08
    & \multicolumn{1}{c}{—}
    & \multicolumn{1}{c}{—}
    & \multicolumn{1}{c}{—}
    & 116.42M  &  7.39 &  45.28 \\
  \textbf{LoRA}
    &   4.46M & 15.95 &  14.49
    &   0.39M & 84.85 &   6.02
    &   1.70M &  7.00 &  45.88 \\
  \textbf{Adapter}
    &   0.33M &  4.10 &  41.28
    &   1.05M & 14.74 &  13.03
    &   0.20M &  2.74 & 137.48 \\
  \bottomrule
\end{tabular}
  } 
\end{table*}

\paragraph{Experimental Setup.}
We systematically evaluate the training efficiency of full fine-tuning, LoRA, and adapter tuning using the COVID variant prediction task from the GUE benchmark \cite{zhou2024dnabert}. 
This task involves classifying sequences into one of nine labels.
We test six models: HyenaDNA-160k, DNABERT-2, Caduceus-131k, Nucleotide Transformer-500M, EVO-1-131k, and GenomeOcean-100M.

For all experiments, we use AdamW with full-precision (FP32) updates, a fixed learning rate of $3\times10^{-5}$, batch size 32, and default model settings unless noted.
For LoRA tuning, we set the rank to $r=8$ and the scaling factor to $\alpha=32$. 
For most models (DNABERT-2, Caduceus, Nucleotide Transformer, GenomeOcean), we apply LoRA to all feed-forward layers.
For HyenaDNA, due to its specialized architecture, we apply LoRA only to the input/output projection layers within the Hyena blocks.
For EVO, given the large size of its feed-forward layers, we target only the key, query, and value matrices in attention blocks.
The adapter tuning keeps all base model parameters frozen and adds a trainable multilayer perceptron adapter with a single hidden layer of size 256.
We conduct all experiments on a single NVIDIA H100 80GB GPU.

\paragraph{Results.}
We present a detailed quantitative comparison of training efficiency across three different fine-tuning strategies in \cref{tab:training-efficiency-part1}.
The table reports the number of trainable parameters, peak GPU memory consumption, and throughput measured in thousands of tokens per second. 
Among all available methods, adapter tuning consistently shows the highest efficiency. 
For example, the DNABERT-2 adapter uses only 0.2 million trainable parameters, consumes 2.09 gigabytes of memory, and reaches a throughput of 124,000 tokens per second. 
In contrast, full fine-tuning for the same model updates 117 million parameters, uses 7.30 gigabytes of memory, and processes only 41,000 tokens per second. 
The adapter module allows users to adjust its internal structure to meet specific hardware or speed constraints.
LoRA also offers strong efficiency gains. 
For DNABERT-2, it reduces the parameter count to 1.49 million and improves throughput to 44,000 tokens per second while reducing memory usage compared to full fine-tuning. 
Full fine-tuning remains the most expensive. 
It requires updating all model parameters, consumes the most GPU memory, and achieves the lowest throughput. 
Due to its extreme cost, we exclude full fine-tuning for the 7B-parameter EVO model.

To assess scalability, we compare LoRA and full fine-tuning across different model sizes in \cref{tab:scaling-efficiency-nt-go}. 
For the Nucleotide Transformer, scaling from 500 million to 2.5 billion parameters lowers the LoRA parameter ratio from 0.9\% to 0.5\%, increases memory savings from 16\% to 32\%, and boosts throughput from 1.20 to 1.39 times compared to full tuning. 
A similar trend appears in GenomeOcean.
Scaling from 100 million to 500 million parameters decreases the parameter ratio from 1.5\% to 0.7\%, increases memory savings from 5\% to 21\%, and raises throughput from 1.01 to 1.27 times. 
These results show that LoRA becomes effective as model size grows, making it a strong choice for adapting large genomic models within \methodac.
These findings align with expectations and demonstrate the functionality of our Model Loader, Model Trainer, and Inference Engine.

\begin{table*}[t] %
  \centering
    \caption{\textbf{Fine-tuning efficiency across different model scales.} 
    We report experimental results to clearly illustrate the scalability of LoRA’s efficiency with Nucleotide Transformer (500M/2.5B) and GenomeOcean (100M/500M) as representative examples. 
    We conduct all experiments on a single NVIDIA H100 (80GB).
     "Full" denotes full-parameter fine-tuning, and "LoRA" denotes low-rank adaptation.}
  \label{tab:scaling-efficiency-nt-go} %
  \resizebox{0.9\textwidth}{!}{%
    \begin{tabular}{llccc} %
      \toprule
      \textbf{Model} & \textbf{Method} & \textbf{Trainable Params} & \textbf{Peak Mem (GB)} & \textbf{Throughput (KTok/s)} \\
      \midrule
      \multirow{2}{*}{\textbf{Nucleotide Transformer-500M}} & \textbf{Full} & 480.45 M & 18.94 & 12.08 \\ %
                       & \textbf{LoRA} & 4.46 M & 15.95 & 14.49 \\ %
      \midrule
      \multirow{2}{*}{\textbf{Nucleotide Transformer-2.5B}} & \textbf{Full} & 2.54 B & 63.95 & 2.40 \\ %
                       & \textbf{LoRA} & 11.86 M & 43.33 & 3.34 \\ %
      \midrule
      \multirow{2}{*}{\textbf{GenomeOcean-100M}} & \textbf{Full} & 116.42 M & 7.39 & 45.28 \\ %
                       & \textbf{LoRA} & 1.70 M & 7.00 & 45.88 \\ %
      \midrule
      \multirow{2}{*}{\textbf{GenomeOcean-500M}} & \textbf{Full} & 534.83 M & 19.58 & 11.85 \\ %
                       & \textbf{LoRA} & 3.97 M & 15.48 & 15.07 \\ %
      \bottomrule
    \end{tabular}%
  }
\end{table*}

\subsection{Benchmarking Different Models}
\label{subsec:Comparison}

We comprehensively benchmark six genomic foundation models with three different fine-tuning strategies across two benchmark suites. 
This experiment highlights the role of the Benchmarker module by comparing model performance and tuning methods on standardized downstream tasks.
This benchmark compares heterogeneous models with different sizes, architectures, pre-training corpora, context lengths, and tokenizers. Thus, the results reflect practical model performance within \methodac, not intrinsic architectural superiority. Controlled architectural conclusions would require matched scale, data, and training budgets.

\paragraph{Experimental Setup.}
We evaluate model performance on tasks from two widely used sources: the GUE benchmark \cite{zhou2024dnabert} and Genomic Benchmarks \cite{grevsova2023genomic}. 
For most benchmark tasks, we report reliable test set performance with the Matthews correlation coefficient.
For the COVID variant prediction task, we follow the benchmark protocol and report the F1 score.
We evaluate the same six models in \cref{subsec:downstream}: HyenaDNA-160k, DNABERT-2,  Caduceus-131k, Nucleotide Transformer-500M, EVO-1-131k, and GenomeOcean-100M.
To provide a robust reference for the computational overhead of foundation models, we compare against several non-foundation-model baselines, including Support Vector Machines and Random Forests trained on 4-mer frequency features, a Convolutional Neural Network operating on one-hot encoded sequences, Position Weight Matrix (PWM)-based motif-scanning pipelines paired with Random Forest or linear SVM classifiers, and Hidden Markov Model (HMM)-based motif-finding approaches.
For each experiment, we consistently use random seeds 14, 28, and 42, and report the mean score along with the standard deviation.

We fine-tune all models with a learning rate of $3.0 \times 10^{-5}$ and the AdamW optimizer in full-precision (FP32) mode. 
For Caduceus, we use a higher learning rate of $1.0 \times 10^{-3}$ to ensure training stability.
For LoRA tuning, we use rank $r=8$ and scaling factor $\alpha=32$. 
Similar to \cref{subsec:downstream}, we apply LoRA to all feed-forward layers in DNABERT-2, Caduceus, Nucleotide Transformer, and GenomeOcean. 
For HyenaDNA, we apply LoRA only to the input and output projection layers within the Hyena blocks. 
For the EVO model, we target the key, query, and value matrices in each attention block, due to the large size of feed-forward layer parameters.
The adapter tuning freezes all base models and appends a trainable multilayer perceptron adapter with a single hidden layer of size 256.
We perform training on a single NVIDIA H100 80GB GPU for most models with a batch size of 32. 
For Nucleotide Transformer-500M, due to its larger memory footprint during full-parameter tuning, we fine-tune it with distributed data parallel across two NVIDIA H100 80GB GPUs.
This also demonstrates the functionality of our distributed training.

\begin{table*}[htbp]
  \centering
  \caption{\textbf{Benchmark across models and tuning methods.} 
We report results on both the GUE benchmark and Genomic Benchmarks. 
We include traditional baselines to establish a performance floor for foundation-model evaluation: a Convolutional Neural Network (CNN), Random Forest, and Support Vector Machine (SVM) classifiers trained on 4-mer frequency features, Position Weight Matrix (PWM)-based motif-scanning pipelines paired with Random Forest or linear SVM classifiers, and Hidden Markov Model (HMM)-based motif-finding methods.
“---” marks settings we did not evaluate due to computational constraints. 
We exclude EVO (7B) from full fine-tuning due to the high computation cost.
Three baselines are trained directly on downstream tasks and thus do not utilize LoRA or Adapter modules.
 ``Full'' denotes full-parameter, ``LoRA'' denotes low-rank adaptation, and ``Adapter'' denotes adapter-based tuning.}
  \label{tab:model-performance-siunitx}
  \resizebox{\textwidth}{!}{%
    \begin{tabular}{
      l l
      r @{\,$\pm$\,} l %
      r @{\,$\pm$\,} l %
      r @{\,$\pm$\,} l %
      r @{\,$\pm$\,} l %
      r @{\,$\pm$\,} l %
      r @{\,$\pm$\,} l %
    }
    \toprule
    \textbf{Model} & \textbf{Size} & \multicolumn{6}{c}{\textbf{GUE}} & \multicolumn{6}{c}{\textbf{Genomic Benchmarks}} \\
    \cmidrule(lr){3-8} \cmidrule(lr){9-14}
      & & \multicolumn{2}{c}{\textbf{Full}} & \multicolumn{2}{c}{\textbf{LoRA}} & \multicolumn{2}{c}{\textbf{Adapter}} & \multicolumn{2}{c}{\textbf{Full}} & \multicolumn{2}{c}{\textbf{LoRA}} & \multicolumn{2}{c}{\textbf{Adapter}} \\
    \midrule
    {\bf HyenaDNA-160k}  &       7M        & 59.91 & 0.22                       & 50.95 & 0.23                           & 25.00 & 0.37                           & 66.71 & 0.18                           & 61.26 & 0.29                           & 36.90 & 0.42 \\
    {\bf  Caduceus-131k}    &    8M        & 50.07 & 1.61                       & 34.70 & 0.19                           & 38.61 & 0.33                           & 65.38 & 0.22                           & 42.10 & 0.36                           & 47.74 & 0.27 \\
    {\bf DNABERT-2}     &        117M        & 65.25 & 0.25                       & 48.39 & 0.11                           & 40.19 & 0.17                           & \bfseries 71.97 & \bfseries 0.20 & 64.58 & 0.25                           & 48.96 & 0.31 \\
    {\bf GenomeOcean-100M}    &    117M      & \bfseries 65.52 & \bfseries 0.57 & \bfseries 59.35 & \bfseries 0.23 & \bfseries 46.65 & \bfseries 0.04 & 68.26 & 0.26 & \bfseries 66.28 & \bfseries 0.19 & \bfseries 53.60 & \bfseries 0.21 \\
    {\bf Nucleotide Transformer-500M} &  480M & 57.63 & 0.26                       & 52.96 & 0.13                           & 32.01 & 0.48                           & 67.99 & 0.24                           & 64.68 & 0.30                           & 43.95 & 0.28 \\
    {\bf EVO-1-131k}         &      7B     & \multicolumn{2}{c}{---}          & 44.97 & 0.79                           & 30.08 & 0.22                           & \multicolumn{2}{c}{---}      & 51.33 & 0.41                           & 33.69 & 0.34 \\
    \midrule
    {\bf CNN} & {---}
    & 45.07 & 1.42
    & \multicolumn{2}{c}{---}
    & \multicolumn{2}{c}{---}
    & 56.58 & 1.50
    & \multicolumn{2}{c}{---}
    & \multicolumn{2}{c}{---} \\
    
    {\bf Random Forest} & {---}
    & 51.79 & 0.12
    & \multicolumn{2}{c}{---}
    & \multicolumn{2}{c}{---}
    & 61.25 & 0.50
    & \multicolumn{2}{c}{---}
    & \multicolumn{2}{c}{---} \\
    
    {\bf Support Vector Machine} & {---}
    & 57.01 & 0.00
    & \multicolumn{2}{c}{---}
    & \multicolumn{2}{c}{---}
    & 59.89 & 0.00
    & \multicolumn{2}{c}{---}
    & \multicolumn{2}{c}{---} \\

    {\bf PWM + Random Forest} & {---}
    & 56.85 & 0.47
    & \multicolumn{2}{c}{---}
    & \multicolumn{2}{c}{---}
    & 62.07 & 0.52
    & \multicolumn{2}{c}{---}
    & \multicolumn{2}{c}{---} \\

    {\bf PWM + Linear SVM} & {---}
    & 55.18 & 0.22
    & \multicolumn{2}{c}{---}
    & \multicolumn{2}{c}{---}
    & 60.14 & 0.38
    & \multicolumn{2}{c}{---}
    & \multicolumn{2}{c}{---} \\

    {\bf Hidden Markov Model} & {---}
    & 50.21 & 0.35
    & \multicolumn{2}{c}{---}
    & \multicolumn{2}{c}{---}
    & 56.92 & 0.61
    & \multicolumn{2}{c}{---}
    & \multicolumn{2}{c}{---} \\
    \bottomrule
    \end{tabular}%
  }
\end{table*}

\paragraph{Results.}
We show results on downstream tasks in \cref{tab:model-performance-siunitx}.
Due to the large size of the EVO model (7B parameters) and its high computational cost, we omit full fine-tuning results for this model. 
We report the averaged Matthews correlation coefficient across all evaluated datasets.
Except for the COVID subset of the GUE benchmark,  we follow the original protocol and use the macro-averaged F1 score. 
The overall GUE score is computed by averaging this F1 value with the correlation coefficients from the remaining tasks.
We observe that GenomeOcean-100M achieves the strongest performance under full fine-tuning on the GUE, while DNABERT-2 performs best on the Genomic Benchmarks. 
Parameter-efficient methods remain competitive. 
For example, LoRA sometimes matches full fine-tuning, as seen with GenomeOcean-100M on Genomic Benchmarks (68.26 vs. 66.28). 
While Adapter method underperforms LoRA in most cases, it narrows the gap and even surpasses LoRA for Caduceus.
This shows that a task-aware adapter improves the task performance.

\begin{figure*}[ht!]
    \centering    \includegraphics[width=0.99\textwidth]{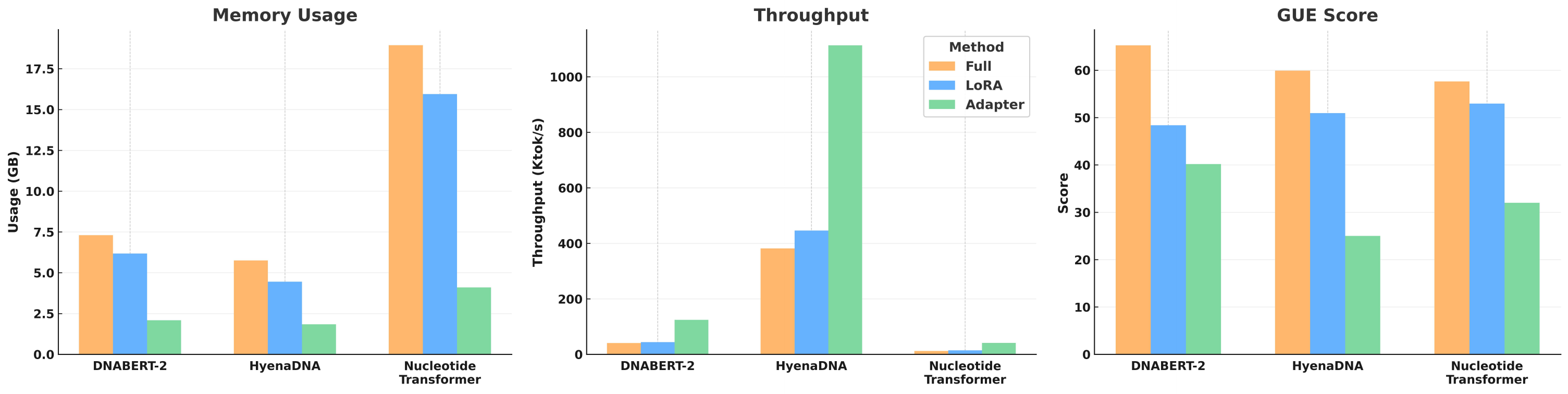}

\caption{\textbf{Trade-off between tuning efficiency and performance.} 
The figure shows memory usage in gigabytes (GB), throughput in kilotokens per second (KTok/s), and averaged scores on the GUE benchmark for three models: DNABERT-2, HyenaDNA-160k, and Nucleotide Transformer-500M. 
We report results for full-tuning (Full), low-rank adaptation (LoRA), and adapter-based fine-tuning (Adapter).
The results highlight the trade-offs between resource efficiency and predictive performance.}
    \label{fig:gfm_performance_comparison}
    \vspace{-1em}
\end{figure*}

We visualize the trade-offs between predictive performance and computational efficiency for DNABERT-2, HyenaDNA, and Nucleotide Transformer in \cref{fig:gfm_performance_comparison} under three fine-tuning strategies: full fine-tuning, low-rank adaptation, and adapter-based methods. 
The figure reports memory usage, training throughput, and GUE scores. 
Adapter tuning reduces memory usage and boosts throughput across all models, though it sacrifices some predictive performance. 
In contrast, full fine-tuning and low-rank adaptation yield higher GUE scores but require more compute. 
This comparison highlights the flexible trade-off space between efficiency and accuracy with \methodac.
Overall, these results demonstrate both the functionality and practical utility of the Benchmarker module.

\paragraph{Analysis of Performance Differences.}
The observed performance gaps likely reflect a combination of model scale, architecture, and pre-training data rather than a single factor. 
Larger models can provide greater representational capacity, but size alone does not explain the results: for example, EVO-1-131k is much larger than the other models but does not consistently achieve the best downstream performance under parameter-efficient tuning. 
Architecture and tokenization also matter because HyenaDNA, Caduceus, DNABERT-2, Nucleotide Transformer, EVO, and GenomeOcean encode sequence context through different mechanisms and token granularities. 
Pre-training data further affects transfer performance, since models trained on different species, genomic regions, or metagenomic corpora may capture different biological signals. 
Therefore, we interpret \cref{tab:model-performance-siunitx} as an empirical comparison of practical model choices within \methodac, rather than a controlled ablation of model architecture.

\subsection{Biological Interpretation of Genomic Models}
\label{subsec:bio_inter}

We demonstrate the capabilities of the Genome Collector and Biological Interpreter modules by analyzing model embeddings through sparse autoencoder-based interpretability.

\paragraph{Experimental Setup.}
We use the Genome Collector to automate genomic data preparation. 
The Data Acquisition component downloads genome sequences for two organisms from NCBI: \textit{Arabidopsis thaliana} and \textit{Bos taurus}. 
We save all downloaded files in a unified directory.
After acquisition, the Data Preprocessing component segments each genome into 1,000 sequences of random length between 500 and 1,000 base pairs, yielding 2,000 total segments. 
We then apply standard quality control procedures, including GC-content correction and removal of N nucleotides. 
Finally, we construct a labeled dataset by assigning each sequence a label corresponding to its sequence length.

To explore representation learning, we train a sparse autoencoder (SAE) with a hidden dimension of 4,096 using model representations and reconstruction losses. 
After training, we use the SAE latent activations to perform regression on the sequence-length labels, allowing us to identify latent features associated with sequence length.

To further validate whether SAE neurons capture biologically meaningful regulatory features, we construct a CTCF-binding dataset from ENCODE K562 ChIP-seq peaks. 
The dataset contains 20,000 IDR peaks as positive examples and length-matched background sequences as negative examples, resulting in 39,024 total sequences. 
We train a logistic regression probe on SAE latent activations to classify CTCF-bound sequences versus background sequences.

\paragraph{Results.}
In the sequence-length analysis, we find that \texttt{latent\_382}, \texttt{latent\_519}, and \texttt{latent\_3519} exhibit strong correlations with sequence length, suggesting that individual SAE latent dimensions encode interpretable sequence-level properties.
In the CTCF-binding analysis, the logistic regression probe achieves an accuracy of 0.7881 for distinguishing CTCF-bound sequences from background sequences. 
We further assess neuron-level importance by individually zeroing the most influential SAE latent dimensions and measuring the resulting drop in classification accuracy. 
Zeroing \texttt{latent\_2939}, \texttt{latent\_3184}, and \texttt{latent\_10} reduces accuracy by 0.0225, 0.0210, and 0.0196, respectively. 
These results show that specific SAE latent dimensions capture biologically meaningful, motif-centric regulatory features.

Together, these analyses demonstrate that the Biological Interpreter can move beyond black-box prediction by linking individual latent features to measurable genomic properties and regulatory signals. 
This provides stronger support for the mechanistic interpretation of genomic foundation models and establishes an open-source framework for connecting model representations with interpretable insights.

\section{Conclusion and Future Work}
\label{sec:Conclusion}We introduce \methodac, the first unified and modular Python framework for streamlining the tuning, deployment, and interpretation of genomic models. 
\methodac~offers an end-to-end pipeline that integrates six key components: Genome Collector for acquiring and preprocessing data; Model Loader for accessing genomic models; Model Trainer for fine-tuning models tailored to specific downstream tasks; Inference Engine for embedding extraction and sequence generation; Benchmarker for evaluating model performance; and Biological Interpreter for interpretability via sparse autoencoders. 
Our experiments demonstrate the utility of \methodac~on key genomic downstream tasks using multiple training methods across diverse models. 
These results highlight its end-to-end usability for real-world analysis. 
\methodac~lowers the technical barrier to using large-scale models in genomics research.
It makes these powerful tools more accessible to the broader research community. 
We discuss the limitations in \cref{sec:limitation}, related works in \cref{sec:related work}, and provide usage examples of \methodac~in \cref{app_sec:use}.

\clearpage

\section*{Acknowledgements}
Weimin Wu would like to thank the anonymous reviewers and program chairs for constructive comments.
JH is partially supported by the Walter P. Murphy Fellowship and the Terminal Year Fellowship (Paul K. Richter Memorial Award) of Northwestern University.
Han Liu is partially supported by NIH R01LM1372201, NSF
AST-2421845, Simons Foundation
MPS-AI-00010513, AbbVie, Dolby, and Chan Zuckerberg Biohub Chicago Spoke Award.
Zhong Wang's work conducted at the U.S. Department of Energy Joint Genome Institute (\url{https://ror.org/04xm1d337}), a DOE Office of Science User Facility, is supported by the Office of Science of the U.S. Department of Energy operated under Contract No. DE-AC02-05CH11231. 
This research was supported in part through the computational resources and staff contributions provided for the Quest high performance computing facility at Northwestern University which is jointly supported by the Office of the Provost, the Office for Research, and Northwestern University Information Technology.
The content is solely the responsibility of the authors and does not necessarily represent the official
views of the funding agencies.

\section*{Impact Statement}

\methodac~provides open-source infrastructure that bridges complex machine learning architectures and practical biological discovery by unifying the genomic model lifecycle, from automated data preprocessing to sparse auto-encoder-based interpretation. By offering zero-code command-line and web interfaces, \methodac~lowers the technical barrier to advanced genomic modeling and enables researchers with varying computational expertise to fine-tune, benchmark, and interpret genomic foundation models. This can accelerate research in personalized medicine, evolutionary biology, and therapeutic design while supporting open, reproducible, and responsible scientific innovation.

At the same time, democratized access to generative sequence models introduces potential dual-use biosecurity risks, including the in silico generation of pathogenic, virulence-associated, or heavily modified viral sequences. Responsible deployment therefore requires safeguards beyond user-level ethical guidelines. Future versions of \methodac~can incorporate software-level guardrails, including sequence screening API integration, pathogen- or virulence-associated motif blocklists, risk-aware generation filters, request logging and auditing, and configurable access controls for high-risk generation tasks. These safeguards can help flag or restrict potentially unsafe outputs while preserving legitimate research use.

\def\arxivfont{\rm}
\bibliographystyle{plainnat}

\bibliography{refs}

\clearpage
\onecolumn
\appendix

\makeatletter
\def\addcontentsline#1#2#3{%
    \addtocontents{#1}{\protect\contentsline{#2}{#3}{\thepage}{\@currentHref}\protected@file@percent}}
\makeatother

\startcontents[sections]

\part*{Appendix}

{
\setlength{\parskip}{.2em}
\printcontents[sections]{ }{1}{\setcounter{tocdepth}{2}}
}

\clearpage

\section{LLM Usage Acknowledgement}
\label{app:llm-usage}

Large language models (LLMs) are used in this project as general-purpose assistive tools to support writing and coding tasks. 
We detail their specific roles below.

\begin{itemize}
\item \textbf{Manuscript Writing:} We use \textbf{GPT-4o} to polish the manuscript, including grammar correction, phrasing refinement, and improving clarity throughout the text.
\item \textbf{Code Development:} We use \textbf{GPT-4o} for debugging, code generation, and resolving implementation issues during the library development.

\end{itemize}

All LLM-assisted content is reviewed and curated by the authors. 
We take full responsibility for the final codebase and manuscript.

\clearpage
\section{Limitations}
\label{sec:limitation}

Although \methodac~provides a unified and practical framework for genomic foundation model training and evaluation, it does not address every advanced research scenario. We identify the following limitations.

\begin{itemize}
    \item \textbf{Scalability of full fine-tuning.}
    Full fine-tuning of very large genomic foundation models remains computationally expensive. For models above several hundred million parameters, such as EVO-7B, full fine-tuning is difficult to perform on a single 80GB GPU. Therefore, \methodac~currently emphasizes parameter-efficient adaptation methods, such as LoRA and adapter tuning, for large-scale models. Future versions will incorporate more memory- and compute-efficient training strategies, including quantization \cite{egiazarian2024extreme}, optimized kernels such as the Liger kernel \cite{hsu2024liger}, and distributed training support.

    \item \textbf{Limitation of context truncation.}
    \methodac~currently truncates sequences that exceed a model's maximum context length while leaving shorter sequences unchanged. This can remove biologically meaningful elements and may bias cross-model comparisons, since short-context models observe less of the original sequence. In our current GUE and Genomic Benchmark experiments, this issue has limited impact because only one task in each benchmark contains sequences longer than 512 bp. However, it becomes important for long-range benchmarks such as DNALONGBENCH \cite{cheng2025dnalongbench}, where sequences can reach 114k bp. Future versions will support sliding-window evaluation and multi-window aggregation to better handle long genomic regions.

    \item \textbf{Limited integration of classical computational biology methods.}
    We conduct experiments using a unified genomic foundation model pipeline and do not yet systematically incorporate traditional computational biology techniques, such as PWM scanning \cite{ambrosini2018pwmscan} and HMM-based motif finding \cite{qin2010hpeak}. These methods remain strong and interpretable baselines for many sequence analysis tasks. Future work will integrate classical methods into the same evaluation framework to enable fair comparison and hybrid modeling.

    \item \textbf{Validation of multi-task learning.}
    \methodac~now supports multi-task learning, but this functionality has not yet been extensively validated across diverse task combinations, label structures, and genomic modalities. In the current version, single-task training remains the default and most thoroughly evaluated setting. Future work will systematically study when multi-task learning improves genomic representation learning and when task interference may reduce performance.

    \item \textbf{Coverage of fine-tuning strategies.}
    We currently include several commonly used fine-tuning methods, but the design space of efficient adaptation remains much broader. Future versions will incorporate additional training strategies and computational acceleration techniques to better support large-scale genomic model development.
\end{itemize}

\clearpage
\section{Related work}
\label{sec:related work}
In the following, we first discuss the genomic foundation models in \cref{subsec: GFM}.
Next, we discuss the existing libraries for natural language models in \cref{subsec: NLP}.

\subsection{Genomic Foundation Models}
\label{subsec: GFM}
Several powerful genomic foundation models have recently emerged to decode the complex language of DNA \cite{avsec2026advancing, brixi2026genome, su2026discrete, zhou2025dnabert, nguyen2024sequence, zhou2024dnabert}.
For the discriminative model domain, 
HyenaDNA \cite{nguyen2023hyenadna} employs implicit convolution layers.
This enables million-token contexts at single-nucleotide resolution to capture complex regulatory interactions and support in-context species classification.
DNABERT-2 \cite{zhou2024dnabert} employs byte pair encoding and a refined transformer architecture.
It improves tasks such as epigenomic mark prediction and transcription factor binding.
Caduceus \cite{schiff2024caduceus} leverages the Mamba architecture \cite{gu2023mamba} with reverse-complement equivariance to improve long-range variant-effect prediction.
Nucleotide Transformer \cite{dalla2025nucleotide} scales to 2.5B parameters with $6$-mer tokenization.
It achieves strong performance in chromatin-feature prediction and functional variant prioritization.
For the generative model domain,
Evo \cite{nguyen2024sequence} introduces a 7B-parameter generative model to extend beyond embedding extraction.
This enables genome-scale predictions across biomolecular modalities.
GenomeOcean \cite{zhou2025genomeocean} further improves generative efficiency for metagenomic sequence synthesis.
It advances applications in synthetic biology.
However, diverse models use a wide range of environments with heterogeneous dependencies and configurations.
This lack of standardization makes it difficult to load, fine-tune, or compare models. 
Such fragmentation increases the technical burden on users and limits the accessibility of genomic models.
To address this, we introduce \methodac, the first Python framework to unify and streamline the end-to-end genomic model workflow.

\subsection{Libraries for Language Models}
\label{subsec: NLP}
In parallel, the language community \cite{comanici2025gemini, guo2025deepseek, wu2025sci2pol, openai2024gpt4technicalreport, touvron2023llamaopenefficientfoundation} has developed numerous frameworks to streamline the adaptation and fine-tuning of language models.
These toolkits target different stages of the language model lifecycle.
For example, LLaMA-Adapter \cite{zhang7llama} improves fine-tuning efficiency, while GPT4All \cite{anand2023gpt4all} enables practical model training and efficient inference on widely available consumer-grade hardware.
Other frameworks address specific training challenges or model architectures: Colossal-AI \cite{li2023colossal} introduces advanced parallelism strategies for efficient large-scale distributed training, FastChat \cite{zheng2023judging} provides specialized tools for building dialogue agents, and Open-Instruct \cite{wang2023far} standardizes methodologies for instruction tuning.
Flexibility, and domain specialization have also emerged as key priorities.
LitGPT \cite{saroufim2025neurips} adopts a modular design to support diverse generative model training paradigms, while LMFlow \cite{diao2024lmflow} helps researchers train language models for specific domains.
Finally, LLaMA-Factory \cite{zheng2024llamafactory} further unifies this ecosystem by integrating multiple efficient fine-tuning techniques into a single, comprehensive toolkit.
However, these frameworks do not translate to the genomic domain.
Genomic models require specialized data formats, biology-informed objectives, domain-specific benchmarks, and meaningful biological insights.
\methodac~fills this gap with tools tailored for genomic models.
It supports data collection, model tuning, inference, benchmarking, and biological interpretation.
Notably, the Biological Interpreter is an open-source tool to decode the hidden internal representations of genomic models.

\clearpage
\section{Experimental Details}
\label{app:traditional_baselines}
In this section, we provide the additional experimental details for \cref{subsec:Comparison}.

\paragraph{Traditional Baseline Models}
We include six baselines: Support Vector Machine (SVM), Random Forest (RF), convolutional neural network (CNN), PWM + RF, PWM + linear SVM, and Hidden Markov Model (HMM). For all baselines, hyperparameters are selected using the validation set only, and the test set is evaluated once using the best validation configuration. All experiments use the same train/validation/test splits as the genomic foundation model experiments.

For SVM and RF, we extract normalized 4-mer frequency features with a stride of 1. Reverse complements are treated as distinct features, and each count vector is normalized by the total number of extracted k-mers. For CNN, we use one-hot encoded DNA sequences as input. For PWM-based baselines, we scan each sequence with position weight matrices and use the resulting motif-score features as input to either RF or linear SVM classifiers. For the HMM baseline, we train an HMM-based motif model and use the resulting sequence likelihood or motif-enrichment scores for downstream classification. Traditional baselines are implemented using \textit{scikit-learn} and \textit{PyTorch}.
See the detailed hyperparameter settings in \cref{tab:traditional_baselines}.

\begin{table}[!ht]
\centering
\small
\caption{\textbf{Hyperparameter settings for traditional baseline models.} Hyperparameters are selected on the validation set, and the test set is evaluated once using the best validation configuration.}
\label{tab:traditional_baselines}
\resizebox{\linewidth}{!}{
\begin{tabular}{p{2.0cm}p{3.2cm}p{6.2cm}p{4.0cm}}
\toprule
\textbf{Model} & \textbf{Input features} & \textbf{Search space} & \textbf{Selected configuration} \\
\midrule
\textbf{SVM} &
4-mer frequency; stride $=1$; reverse complements treated separately; normalized by total k-mer count &
Kernel $\in \{\text{linear}, \text{rbf}\}$; $C \in \{0.1, 1, 10, 100\}$; for RBF, $\gamma \in \{\text{scale}, 10^{-3}, 10^{-2}, 10^{-1}\}$ &
Kernel $=\text{rbf}$; $C=10$; $\gamma=\text{scale}$ \\

\midrule

\textbf{RF} &
Same 4-mer frequency features as SVM &
Trees $\in \{100,300,500\}$; max depth $\in \{\text{None},10,20,40\}$; min samples split $\in \{2,5,10\}$; min samples leaf $\in \{1,2,4\}$; max features $\in \{\text{sqrt}, \text{log2}, 0.5\}$ &
Trees $=500$; max depth $=20$; min samples split $=2$; min samples leaf $=1$; max features $=\text{sqrt}$ \\

\midrule

\textbf{CNN} &
One-hot encoded DNA sequence &
3 convolutional layers; channels $(128,256,256)$; kernel sizes $(7,5,3)$; ReLU; global max pooling; FC hidden dim $=256$; dropout $\in \{0.3,0.5\}$; learning rate $\in \{10^{-4},5\times10^{-4},10^{-3}\}$; batch size $\in \{32,64,128\}$; weight decay $\in \{0,10^{-5},10^{-4}\}$ &
Adam; learning rate $=10^{-3}$; batch size $=64$; dropout $=0.5$; weight decay $=10^{-5}$; max epochs $=30$; early stopping patience $=5$ \\

\midrule

\textbf{PWM + RF} &
PWM motif-scanning scores &
Same RF search space as above &
Trees $=500$; max depth $=20$; min samples split $=2$; min samples leaf $=1$; max features $=\text{sqrt}$ \\

\midrule

\textbf{PWM + SVM} &
PWM motif-scanning scores &
Linear SVM with $C \in \{0.1,1,10,100\}$ &
Linear kernel; $C=10$ \\

\midrule

\textbf{HMM} &
HMM-derived sequence likelihood &
Number of hidden states $\in \{2,4,8\}$; covariance/transition smoothing $\in \{10^{-4},10^{-3},10^{-2}\}$ &
Hidden states $=4$; smoothing $=10^{-3}$ \\
\bottomrule
\end{tabular}
}
\end{table}

\clearpage
\section{Details of Supported Models}
The \cref{tab:model_overview_placeholder} show \methodac~'s list of supported models.
\begin{table}[htbp] %
    \centering %
    \caption{{\bf Supported models in \methodac~with their available variants.} Models vary by either parameter size (e.g., GenomeOcean: 100M/500M/4B) or input sequence length (e.g., Hyenadna: 1K to 1M). "Variant Type" specifies the axis of variation, and "Variants" lists the available options.}
    \label{tab:model_overview_placeholder}
    \resizebox{0.7\textwidth}{!}{
    \begin{tabular}{lcccc}
        \toprule
        \textbf{Model} & \textbf{Variant Type} & \textbf{Variants} \\
        \midrule
        \textbf{Hyenadna} & Sequence Length & 1K/16K/32K/160K/450K/1M & & \\
        \textbf{DNABERT-2} & Parameter Size & 117M & & \\
        \textbf{Caduceus} & Sequence Length & 1K/131K & & \\
        \textbf{Nucleotide Transformer} & Parameter Size & 50M/100M/250M/500M/1B/2.5B & & \\
        \textbf{EVO} & Sequence Length & 8K/131K & & \\
        \textbf{GenomeOcean} & Parameter Size & 100M/500M/4B & & \\
        \bottomrule
    \end{tabular}
    }
\end{table}

\clearpage
\section{Additional Benchmarking on Broader Genomic Tasks}
\label{app:additional_benchmark}

To further assess the generalizability of \textsc{Genome-Factory} beyond the benchmark suites reported in \cref{subsec:Comparison}, we conduct additional experiments on three complementary settings: (i) epigenetic mark prediction with larger genomic foundation models, (ii) unsupervised multi-species representation evaluation, and (iii) genomic sequence-to-function and long-range benchmarks. These experiments further validate the extensibility of the Benchmarker module and provide a broader comparison with recent genomic models and benchmark suites.

\paragraph{GUE Epigenetic Marks with Large Models.}
We first evaluate larger genomic foundation models on the 10 epigenetic mark prediction tasks from the Genome Understanding Evaluation (GUE) benchmark \cite{zhou2024dnabert}. 
We compare GenomeOcean-4B and EVO 2-7B using LoRA fine-tuning and report the average Matthews correlation coefficient (MCC) across all 10 tasks. 
GenomeOcean-4B achieves an average MCC of 60.43, while EVO 2-7B achieves 61.77. 
These results show that GenomeOcean-4B remains competitive with a substantially larger state-of-the-art generative genomic model under the same parameter-efficient fine-tuning protocol.

\paragraph{Unsupervised Multi-species Representation Evaluation.}
To evaluate whether model embeddings capture biologically meaningful species-level structure without supervised labels, we extract mean-pooled sequence embeddings on the Genomic Benchmarks species classification task \cite{grevsova2023genomic}. 
We then apply $k$-means clustering to the embeddings and measure the adjusted Rand index (ARI) against the ground-truth species labels. 
As shown in \cref{tab:species_ari}, GenomeOcean-4B achieves the strongest unsupervised clustering performance, nearly doubling the next-best model. 
This suggests that its learned representations capture richer multi-species biological structure without task-specific supervised training.

\begin{table}[!ht]
\centering
\caption{\textbf{Unsupervised representation quality on the Genomic Benchmarks species classification task.} We extract mean-pooled embeddings, apply $k$-means clustering, and report the adjusted Rand index (ARI) against ground-truth species labels.}
\label{tab:species_ari}
\begin{tabular}{lcccc}
\toprule
\textbf{Model} & \textbf{DNABERT-2} & \textbf{HyenaDNA-160k} & \textbf{EVO 2-7B} & \textbf{GenomeOcean-4B} \\
\midrule
\textbf{ARI} & 0.167 & 0.146 & 0.082 & 0.321 \\
\bottomrule
\end{tabular}
\end{table}

\paragraph{Evaluation on GUANinE and Long-range Genomic Benchmarks.}
We further expand the evaluation to include GROVER and two additional benchmark suites: GUANinE v1.0 \cite{ioannidis2024guanine} and the Genomics Long-Range Benchmark (LRB) \cite{trop2024genomics}. 
GUANinE provides sequence-to-function tasks for genomic AI models, while LRB emphasizes long-range genomic modeling. 
We evaluate all models using LoRA fine-tuning and report MCC. 
\cref{tab:additional_benchmarks} summarizes the results across GUE, Genomic Benchmarks, GUANinE, and LRB.

\begin{table}[!ht]
\centering
\caption{\textbf{Additional benchmark results with broader benchmark suites.} We report MCC under LoRA fine-tuning. GUE denotes the Genome Understanding Evaluation benchmark, GB denotes Genomic Benchmarks, GUANinE denotes GUANinE v1.0, and LRB denotes the Genomics Long-Range Benchmark.}
\label{tab:additional_benchmarks}
\begin{tabular}{lcccc}
\toprule
\textbf{Model} & \textbf{GUE} & \textbf{GB} & \textbf{GUANinE} & \textbf{LRB} \\
\midrule
\textbf{GROVER} & ${\bf 61.38 \pm 0.33}$ & $65.42 \pm 0.27$ & $58.15 \pm 0.41$ & $42.73 \pm 0.55$ \\
\textbf{DNABERT-2} & $48.39 \pm 0.11$ & $64.58 \pm 0.25$ & $62.80 \pm 0.35$ & $45.19 \pm 0.48$ \\
\textbf{HyenaDNA-160k} & $50.95 \pm 0.23$ & $61.26 \pm 0.29$ & $55.47 \pm 0.39$ & ${\bf 51.36 \pm 0.62}$ \\
\textbf{GenomeOcean-100M} & $59.35 \pm 0.23$ & ${\bf 66.28 \pm 0.19}$ & ${\bf 63.14 \pm 0.28}$ & $48.92 \pm 0.51$ \\
\bottomrule
\end{tabular}
\end{table}
 
On GUANinE, GenomeOcean-100M obtains the best performance, suggesting strong sequence-to-function adaptation under LoRA fine-tuning. 
On LRB, HyenaDNA achieves the best result, likely because its long-context architecture is better suited to long-range genomic dependencies. 
Together, these results show that \textsc{Genome-Factory} can support broader benchmark coverage and more diverse evaluation settings beyond the original GUE and Genomic Benchmarks experiments.

\clearpage
\section{Usage Examples of \methodac}
\label{app_sec:use}

We provide usage examples of \methodac~through two interfaces: the command-line interface (CLI) and the web-based interface (WebUI). 
We begin with detailed CLI usage examples, including (i) data downloading in \cref{app_subsec:dl}, (ii) data processing in \cref{app_subsec:dp}, (iii) model training in \cref{app_subsec:mt}, (iv) model inference in \cref{app_subsec:mi}, and (v) biological interpretation in \cref{app_subsec:bi}.
Then we conclude with the WebUI demonstration in \cref{app_sec:webui}. 

Genome-Factory relies on YAML configuration files to define tasks, with example files available in \textit{``genomeFactory/Examples''}. 
Users can customize the parameters in these files as long as the required YAML structure is preserved.

\subsection{Data Downloading}
\label{app_subsec:dl}

Users can download data from NCBI.
It supports downloading from a config file or an interface.

\paragraph{Using a Config File.} 
Specify download parameters in a YAML file. 
It supports both species-based and link-based downloads.

\begin{itemize}
    \item Download by species.

\begin{lstlisting}[aboveskip=2pt, belowskip=2pt, backgroundcolor=\color{codegray}]
genomefactory-cli download genomeFactory/Examples/download_by_species.yaml
\end{lstlisting}

\item Download by Link.
\begin{lstlisting}[aboveskip=2pt, belowskip=2pt, backgroundcolor=\color{codegray}]
genomefactory-cli download genomeFactory/Examples/download_by_link.yaml
\end{lstlisting}
\end{itemize}

\paragraph{Using the Interactive Interface.}
Run the command without a config file and follow the prompts to specify your download criteria (this supports both species-based and link-based downloads).
\begin{lstlisting}[aboveskip=2pt, belowskip=2pt, backgroundcolor=\color{codegray}]
genomefactory-cli download
\end{lstlisting}

For both two methods, the list of species and their taxonomy IDs used for downloads is stored in
\textit{``genomeFactory/Data/Download/Datasets\_species\_taxonid\_dict.json''}. 
Users can extend it by adding new species–taxonomy ID pairs to enable downloads for additional species.

\subsection{Data Processing}
\label{app_subsec:dp}

\methodac~provides tools to prepare data for model fine-tuning. 
This includes (i) processing data downloaded from NCBI (\cref{app_subsubsec:pnd}), (ii) formatting users' own custom datasets (\cref{app_subsubsec:pcd}), and (iii) processing specialized genomic data (\cref{app_subsubsec:sp}).

\subsubsection{Processing NCBI Data}
\label{app_subsubsec:pnd}

Follow the steps below to process the data downloaded from NCBI.
\begin{itemize} 
  \item Gather data downloaded into a single folder (steps in \cref{app_subsec:dl} finish this automatically).
  
  \item Run the processing command with a config file.
\begin{lstlisting}[aboveskip=2pt, belowskip=2pt, backgroundcolor=\color{codegray}]
genomefactory-cli process genomeFactory/Examples/process_normal.yaml
\end{lstlisting}

  \item The processed data will be ready for model fine-tuning.
\end{itemize}

\subsubsection{Preparing Custom Datasets}
\label{app_subsubsec:pcd}

Follow the steps below to process the custom data.

\begin{itemize}
  \item Separate the data into three CSV files: \textit{``train.csv''}, \textit{``dev.csv''}, and \textit{``test.csv''}.
  \item Each CSV file must have two columns:
    (i) The first column should contain the DNA sequences (e.g., \textit{``sequence''}).
    (ii) The second column should contain the corresponding labels (e.g., \textit{``label''}).
    For classification tasks, labels should be integers (e.g., 0, 1, 2).
    For regression tasks, labels should be continuous numbers.
        
  \item Place these three CSV files (\textit{``train.csv''}, \textit{``dev.csv''}, and \textit{``test.csv''}) together in a single folder.
  \item Specify this folder as the input data directory in your training configuration YAML file.
\end{itemize}

\subsubsection{Specialized Processing}
\label{app_subsubsec:sp}

\methodac~provides specialized dataset generation tools for common genomic tasks.

\begin{itemize}
  \item Promoter region dataset: Generate promoter vs. non-promoter classification data for human, mouse, and zebrafish genomes.
\begin{lstlisting}[aboveskip=2pt, belowskip=2pt, backgroundcolor=\color{codegray}]
genomefactory-cli process genomeFactory/Examples/process_promoter.yaml
\end{lstlisting}

  \item Epigenetic mark dataset: Create gene body sequences with H3K36me3 signal classification for human and mouse genomes.
\begin{lstlisting}[aboveskip=2pt, belowskip=2pt, backgroundcolor=\color{codegray}]
genomefactory-cli process genomeFactory/Examples/process_emp.yaml
\end{lstlisting}

  \item Enhancer region dataset: Build enhancer vs. non-enhancer classification data for human and mouse genomes.
\begin{lstlisting}[aboveskip=2pt, belowskip=2pt, backgroundcolor=\color{codegray}]
genomefactory-cli process genomeFactory/Examples/process_enhancer.yaml
\end{lstlisting}
\end{itemize}
All three datasets feature quality control, configurable train/val/test splits, and output CSV files with \textit{``sequence'', ``label''} format.

\subsection{Model Training}
\label{app_subsec:mt}

To fine-tune GFMs, \methodac~supports two primary task types: classification and regression.
Users can specify the desired \textit{task\_type} in the training YAML configuration file.

Follow the steps below to fine-tune GFMs using three different methods.

\paragraph{Full-parameter Fine-tuning.} 
The following command applies to all models except for EVO.
We do not support the full-parameter fine-tuning for EVO now.

\begin{lstlisting}[aboveskip=2pt, belowskip=2pt, backgroundcolor=\color{codegray}]
genomefactory-cli train genomeFactory/Examples/train_full.yaml
\end{lstlisting}

\paragraph{Low-rank Adaptation Fine-tuning (LoRA).} 

\begin{itemize}
    \item 
For general models (except for EVO), users can use:
\begin{lstlisting}[aboveskip=2pt, belowskip=2pt, backgroundcolor=\color{codegray}]
genomefactory-cli train genomeFactory/Examples/train_lora.yaml
\end{lstlisting}
\item 
For the EVO model, users need to use:
\begin{lstlisting}[aboveskip=2pt, belowskip=2pt, backgroundcolor=\color{codegray}]
genomefactory-cli train genomeFactory/Examples/train_evo_lora.yaml
\end{lstlisting}
\end{itemize}

For all models, users can specify target modules in the YAML file: (i) \textit{``all''} - targets all linear layers;
(ii) \textit{``all\_in\_and\_out\_proj''} - targets input/output projection linear layers and the final classification layer;
and (iii) \textit{``custom''} - specifies module names by users.

\paragraph{Adapter-based Fine-tuning.} The following command applies to all models.

\begin{lstlisting}[aboveskip=2pt, belowskip=2pt, backgroundcolor=\color{codegray}]
genomefactory-cli train genomeFactory/Examples/train_adapter.yaml
\end{lstlisting}

In \textit{``genomeFactory/Train/workflow/adapter/adapter\_model/Adapter.py''}, users can customize the adapter architecture for better performance on specific downstream tasks.

For all three methods, training settings (e.g., batch size, learning rate, and training epochs) can be customized in the respective YAML files.

\textbf{Benchmarking:} After fine-tuning, performance metrics are saved to a JSON file. 
Users can use these metrics for benchmarking (e.g., comparing the performance of different models or tuning methods).

\subsection{Model Inference}
\label{app_subsec:mi}

Follow the steps below to use genomic models for embedding extraction and generation.

\paragraph{Embedding Extraction.} Extract the last hidden state embeddings from sequences.

\begin{itemize}
\item 
  For the general case (except for EVO):
\begin{lstlisting}[aboveskip=2pt, belowskip=2pt, backgroundcolor=\color{codegray}]
genomefactory-cli inference genomeFactory/Examples/inference_extract.yaml
\end{lstlisting}

     \item    For EVO:
\begin{lstlisting}[aboveskip=2pt, belowskip=2pt, backgroundcolor=\color{codegray}]
genomefactory-cli inference genomeFactory/Examples/inference_extract_evo.yaml
\end{lstlisting}
\end{itemize}

\paragraph{Generation.} Generate new DNA sequences based on prompts. 
  Applicable to generative models.

\begin{itemize}
\item 
  For GenomeOcean:
\begin{lstlisting}[aboveskip=2pt, belowskip=2pt, backgroundcolor=\color{codegray}]
genomefactory-cli inference genomeFactory/Examples/inference_generation_genomeocean.yaml
\end{lstlisting}

   \item 
  For EVO:
\begin{lstlisting}[aboveskip=2pt, belowskip=2pt, backgroundcolor=\color{codegray}]
genomefactory-cli inference genomeFactory/Examples/inference_generation_evo.yaml
\end{lstlisting}

\end{itemize}

\subsection{Biological Interpretation}
\label{app_subsec:bi}

\methodac~provides comprehensive tools for understanding and interpreting GFMs through a sparse auto-encoder (SAE) to provide biological insights into model behavior.

Follow the steps below to implement the workflow for interpreting.

\paragraph{Train an SAE Model.}
Use the following command to train SAE with the DNA sequences.

\begin{lstlisting}[aboveskip=2pt, belowskip=2pt, backgroundcolor=\color{codegray}]
genomefactory-cli sae_train genomeFactory/Examples/sae_train.yaml
\end{lstlisting}

Configure the following parameters in the YAML file.
Users need to specify the DNA sequence file as \textit{data\_file}.
Use \textit{model\_name} to assign a name or checkpoint path for the genomic model.
Set \textit{d\_model} to the output dimension of the genomic model used to generate the embeddings.
Define the SAE bottleneck dimension using \textit{d\_hidden}.
It controls the number of sparse latent features.

\begin{lstlisting}[aboveskip=2pt, belowskip=2pt, backgroundcolor=\color{codegray}]
data_file: "<YOUR_DNA_SEQUENCE_FILE>"
d_model: "<GENOMIC_MODEL_OUTPUT_DIMENSION>"
d_hidden: "<SAE_HIDDEN_DIMENSION>"
model_name: "<MODEL_NAME_AND_PATH>"
\end{lstlisting}

\paragraph{Downstream Evaluations with Ridge Regression.}

Users need to generate hidden embeddings from genomic models and save the results as a CSV file at the path specified by \textit{``<GENOMIC\_MODEL\_FEATURE\_CSV\_PATH>''} (by \cref{app_subsec:mi}).
Then use [CLS] tokens embedding or mean-pooling embedding to do further analysis.

\begin{itemize}
\item 
Use the [CLS] token latent embedding for analysis.
\begin{lstlisting}[aboveskip=2pt, belowskip=2pt, backgroundcolor=\color{codegray}]
genomefactory-cli sae_regression genomeFactory/Examples/sae_regression.yaml
\end{lstlisting}

Configure the following parameters in the YAML file.
Users need to specify the DNA sequence embedding file as \textit{csv\_path}.
Use \textit{sae\_checkpoint\_path} to assign the checkpoint path of the SAE.

\begin{lstlisting}[aboveskip=2pt, belowskip=2pt, backgroundcolor=\color{codegray}]
csv_path: "<GENOMIC_MODEL_FEATURE_CSV_PATH>"
sae_checkpoint_path: "<SAE_CHECKPOINT_PATH>"
type: "<first_token>"
\end{lstlisting}

\item 
Use the mean-pooling latent embedding for analysis.
\begin{lstlisting}[aboveskip=2pt, belowskip=2pt, backgroundcolor=\color{codegray}]
genomefactory-cli sae_regression genomeFactory/Examples/sae_regression.yaml
\end{lstlisting}

Configure the following parameters in the YAML file.
\begin{lstlisting}[aboveskip=2pt, belowskip=2pt, backgroundcolor=\color{codegray}]
csv_path: "<GENOMIC_MODEL_FEATURE_CSV_PATH>"
sae_checkpoint_path: "<SAE_CHECKPOINT_PATH>"
type: "<mean>"
\end{lstlisting} 
\end{itemize}

\subsection{Usage via WebUI}
\label{app_sec:webui}

Use the following command to access all \methodac~functionalities by a graphical interface.
\begin{lstlisting}[aboveskip=2pt, belowskip=2pt, backgroundcolor=\color{codegray}]
genomefactory-cli webui
\end{lstlisting}
This command launches a web server. Open the provided URL in users' browsers to use the WebUI.

\stopcontents[sections]

\end{document}